\documentclass[aps,prd,groupedaddress,twocolumn,floatfix]{revtex4-2}
\usepackage{amsmath}
\usepackage{amssymb}
\usepackage{graphicx}

\begin{document}

\title{Eigenmodes of the Laplacian on Hyperbolic Lattices}
\author{Eric Petermann}
\email{eric-petermann@web.de}
\author{Haye Hinrichsen}
\email{haye.hinrichsen@uni-wuerzburg.de}
\affiliation{Faculty for Physics and Astronomy, Julius Maximilians University W\"urzburg, Am Hubland, 97074 Würzburg, Germany}

\date{\today}

\begin{abstract}
We examine a specific category of eigenfunctions of the lattice Laplacian on $\{p,q\}$-tessellations of the Poincaré disk that bear resemblance to plane waves in the continuum case. Our investigation reveals that the lattice eigenmodes deviate from the continuum solutions by a factor that depends solely on the local inclination of the vertex in relation to the wave's propagation direction. This allows us to compute certain eigenfunctions by numerical and analytical methods. For various special cases we find explicit exact eigenfunctions and their eigenvalues on the infinite lattice.
\end{abstract}

\maketitle

%=====================================================================
% Special definitions
\def\vec#1{\mathbf{#1}}
\parskip 2mm
%=====================================================================

%=====================================================================
\section{Introduction}
%=====================================================================
Formulating a consistent theory of quantum gravity remains one of the greatest challenges in modern physics. Such a theory is expected to not only describe the curvature of spacetime but also account for its potentially discrete nature at the Planck scale. To gain a deeper understanding of the interaction between curvature and discretization, quantum systems residing on tessellations of hyperbolic spaces with constant negative curvature~\cite{coxeter1957crystal} have emerged as a highly promising testing ground. In this context, various innovative experimental implementations, in particular photonic crystals~\cite{ozawa2019topological,kollar2019hyperbolic,yu2020topological} and electronic circuits on hyperbolic lattices~\cite{lee2018topolectrical,helbig2020generalized,dong2021topolectric,lenggenhager2021electric,lenggenhager2022simulating,chen2023hyperbolic,gluscevich2023dynamic}, are currently attracting significant attention.

These groundbreaking experiments have rekindled the interest in the fascinating mathematical properties of discretized hyperbolic spaces, particularly isometric tilings of the Poincaré disk. A significant milestone in this realm has been the development of hyperbolic band theory, drawing inspiration from solid-state physics~\cite{maciejko2021hyperbolic,boettcher2022crystallography,maciejko2022automorphic,ikeda2021hyperbolic,attar2022selberg,bzduvsek2022flat,mosseri2022aharonov,cheng2022band,zhang2023hyperbolic,liu2023higher,kollar2020line}. This theory combines the concept of Bloch waves~\cite{ashcroft2022solid} with group-theoretical ideas~\cite{katok1992fuchsian}, allowing one to describe specific categories of wave functions that exist within hyperbolic crystals.

From a mathematical point of view, the challenge of describing the band structure is closely intertwined with understanding the spectral properties of the discretized Laplace-Beltrami operator~\cite{boettcher2020quantum,bienias2022circuit}. However, unlike the Euclidean case, where the eigenfunctions and eigenvalues of the lattice Laplacian can be easily computed, the eigenfunctions on hyperbolic lattices are significantly more intricate and generally cannot be expressed using closed formulas. The current band theory partially addresses this issue by assuming the existence of a translation subgroup on the lattice, which imposes certain limitations on possible lattice geometries. Nevertheless, a comprehensive classification of the eigenfunctions and eigenvalues of the hyperbolic lattice Laplacian remains elusive. Solving this problem would not only be valuable in the context of band theory but also hold profound implications for other fields, ranging from the AdS/CFT correspondence~\cite{maldacena1999large,witten1998anti,ammon2015gauge,boyle2020conformal,asaduzzaman2020holography,brower2021lattice,basteiro2022towards} to applications in quantum information theory~\cite{ryu2006holographic,vidal2007entanglement,pastawski2015holographic,jahn2021holographic}.

As a further advancement in this direction, we investigate here a specific category of eigenfunctions of the Laplacian on infinite $\{p,q\}$-tilings of the Poincaré disk. These eigenfunctions on the lattice bear resemblance to plane waves in the continuum. To accomplish this, we introduce a novel geometric concept, referred to as the \textit{inclination} of a vertex. This concept plays a pivotal role in enabling us to provide exact solutions for a subset of points within the parameter space. Additionally, we propose two complementary numerical methods that facilitate the efficient computation of the eigenfunctions and their corresponding eigenvalues on the infinite lattice.

The article is organized as follows. In the following Section we first summarize relevant aspects of the continuum theory, discussing the spectral properties of the Laplace-Beltrami operator on the Poincaré disk. In Sect.~\ref{sec:LatticeLaplacian} we review isometric tilings of the Poincaré disk as well as the question of how to define an appropriate discretized version of the Laplace-Beltrami operator on such lattices. We then demonstrate that there are special choices of the parameters for which the eigenfunctions on the lattice coincide with those in the continuum.

In order to arrive at a general solution, we then introduce the concept of \textit{inclination} in Sect.~\ref{sec:Inclination}. The central hypothesis of this work is that the eigenfunctions on the lattice differ only slightly from the eigenfunctions in the continuum and that the corresponding correction function depends exclusively on the inclination. In Sect.~\ref{sec:CorrectionFunction}, a linear difference equation is then derived for this correction function, which can also be used to calculate the solution numerically through iteration. Using a Fourier transformation, we then arrive at an eigenvalue problem for the Fourier coefficients in Sect.~\ref{sec:FourierAnalysis}. In certain points of the parameter space, we finally show that the corresponding matrix exhibits a block structure, allowing us to derive a various exact solutions. The article ends with concluding remarks in Sect.~\ref{sec:Conclusions}. Technical details are given in the appendices A-D.
\vfill

%=====================================================================
\section{Continuum Theory}
\label{sec:Continuum}
%=====================================================================

%---------------------------------------------------------------------
\subsection{The Poincaré Disk}
%---------------------------------------------------------------------
%
Before we begin, let us briefly summarize what is known about the continuum theory. The Poincaré disk serves as a conformal representation of the hyperbolic plane $H^2$ with constant negative curvature. It is defined on an open unit disk $\mathbb D \subset \mathbb C$ equipped with the conformal metric
\begin{equation}
\label{Metric}
\text{d} s^2 \;=\; \frac{\text{d} z \text{d} \overline z}{(1-z \overline z)^2}\,.
\end{equation}
This metric is preserved under the isometries
\begin{equation}
\label{Isometry}
z \;\mapsto\; w(z)=e^{i\eta}\frac{a-z}{1-z \overline a}\,\,,
\end{equation}
where $\eta \in[0,2\pi)$ and $a\in\mathbb D$.

On the Poincaré disk the geodesic lines are represented as segments of circles. The geodesic distance between two points is given by
\begin{equation}
\label{GeodesicDistance}
\text{d}(z,z') \;=\; \frac12\, \text{arcosh}\Bigl( 1+\frac{2|z-z'|^2}{(1-|z|^2)(1-|z'|^2)} \Bigr)
\end{equation}
and is preserved under the isometries~(\ref{Isometry}).

With the metric (\ref{Metric}) the Laplace-Beltrami operator on the Poincaré disk is given by
\begin{equation}
\label{LaplaceBeltrami}
\Delta_g \;=\; \frac{1}{\sqrt{g}}\partial_i\Bigl( \sqrt{g}\,g^{ij}\partial_j \Bigr) \;=\;4(1-z\overline z)^2\partial_z \partial_{\overline z}\,.
\end{equation}

%---------------------------------------------------------------------
\subsection{Construction of Plane-Wave Eigenmodes}
%---------------------------------------------------------------------

To comprehend the analogue of plane waves on the Poincaré disk in the continuum case, let us first recall the situation in the Euclidean $\mathbb R^n$. Here the plane-wave solutions of the eigenvalue problem $\Delta \psi = -\lambda \psi$ are given by $\psi_{\vec k}(\vec x) = e^{i \vec k \cdot \vec x}$, where $\vec k \in \mathbb R^n$ denotes the momentum vector and $\lambda = k^2$ is the corresponding eigenvalue. Using the Euclidean distance $d(\vec x,\vec y)=||\vec x-\vec y||$, these eigenfunctions can be expressed in terms of the limit
\begin{equation}
\label{EuclideanPlaneWaveLimit}
\psi_{\vec k}(\vec x) =
e^{i \vec k \cdot \vec x} =
\lim_{r \to \infty} e^{-ik\bigl( d(\vec x, r \vec k)-d(0, r \vec k) \bigr)} \,,
\end{equation}
where $r \vec k$ is a point approaching infinity in the direction of~$\vec k$. This elucidates that a plane wave can be interpreted as a radial wave originating from a point at infinite distance in a specific direction.

We can now apply the same rationale to construct the analogue of plane waves on the Poincaré disk. The only requirement is to replace the Euclidean distance $d(\vec x,\vec y)$ by the geodesic distance $\text{d}(z,z')$ defined in (\ref{GeodesicDistance}). Note that points at infinity are now residing on the boundary of the disk $\partial \mathbb D$ where $|z|=1$. This suggests to define the hyperbolic equivalent of a plane wave on the Poincaré disk in analogy to (\ref{EuclideanPlaneWaveLimit}) as
\begin{equation}
\label{psilimit}
\psi_{\kappa,b}(z) \;:=\; \lim_{r \to 1} e^{-i\kappa\bigl( d(z,rb)-d(0,rb) \bigr)}\,.
\end{equation}
Here, $b = e^{i\beta}\in\partial\mathbb D$ is a point on the boundary of the disk, which we refer to as the \textit{source} of the plane wave in our study. Expanding the argument of the exponential function in (\ref{psilimit}) around $r=1$ leads to
\begin{equation}
\lim_{r \to 1}\bigl( d(z,rb)-d(0,rb) \bigr)\;=\;\frac12 \ln \frac{\bigl|z-b\bigr|^2}{1-|z|^2}
\end{equation}
and consequently,
\begin{equation}
\psi_{\kappa,b}(z) \;=\; \Bigl( \frac{1-|z|^2}{|1-z \overline b|^2} \Bigr)^{i\kappa/2}\,.
\end{equation}
\begin{figure}
\includegraphics[width=85mm]{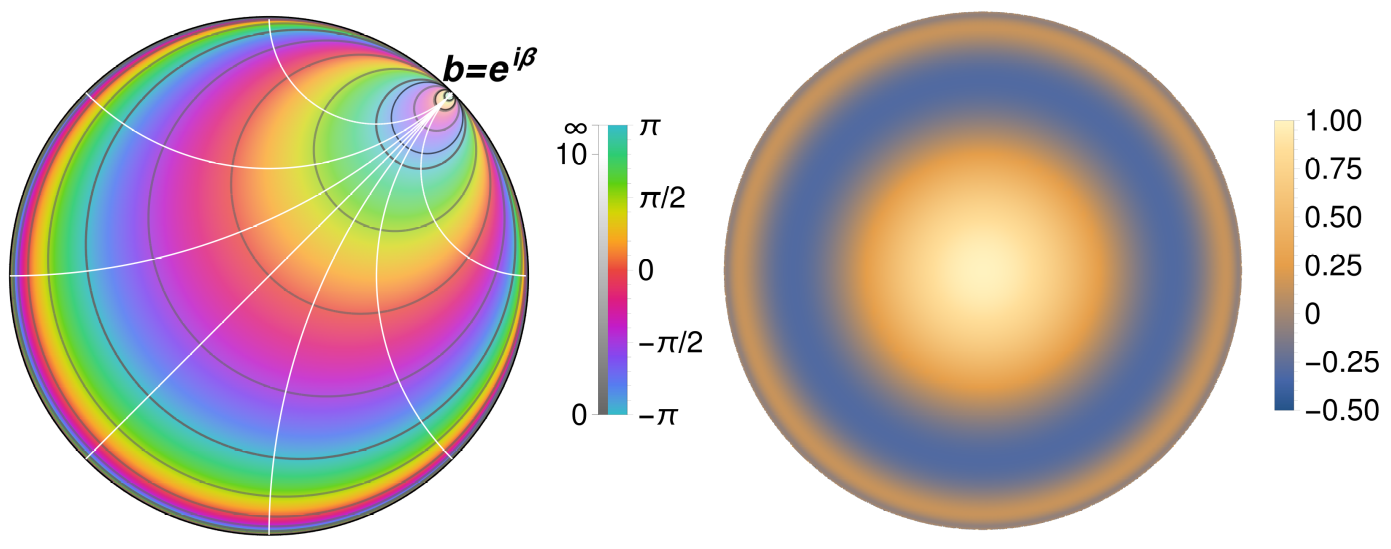}
\caption{Continuum eigenfunctions on the Poincaré disk. Left: Plane wave eigenfunction $\psi_{k,b}(z)$ for $b=e^{i \pi/4}$ and $k=4$. The geodesics and horocycles are depicted as white and black circles, respectively. Right: The corresponding radial eigenfunction $u_{k}^m(z)$ for $m=0$.
\label{fig:Eigensolutions}}
\end{figure}
By applying the Laplace-Beltrami operator (\ref{LaplaceBeltrami}), it can be verified that this function solves the eigenvalue problem $\Delta_g \psi_{\kappa,b}=-\lambda \psi_{\kappa,b}$. However, the corresponding eigenvalue $\lambda=\kappa(\kappa+2i)$ turns out to be complex-valued. To acquire a real positive spectrum with normalizable oscillatory eigenfunctions as in the Euclidean case, it is therefore useful to redefine the parameter $\kappa$ as $k=\kappa + i$, enabling us to interpret $k$ as a generalized momentum. This notation leads to the commonly accepted definition of hyperbolic plane waves in the literature~\cite{boettcher2020quantum}, specifically,
\begin{equation}
\label{planewave}
\psi_{k,b}(z) = \Bigl( \frac{1-|z|^2}{|1-z e^{-i\beta}|^2} \Bigr)^{\frac12(1+ik)}\,,
\end{equation}
which corresponds to the eigenvalue
\begin{equation}
\lambda=1+k^2\,.
\end{equation}
Note that these eigenfunctions differ from Euclidean plane waves in that they exhibit an additional exponential damping as well as a constant offset in the eigenvalue, which is a consequence of the constant negative curvature of the underlying hyperbolic space.

A typical wave function of the form~(\ref{planewave}) is shown in the left panel of Fig.~\ref{fig:Eigensolutions}. As can be seen, the wave emanates from a point located at the boundary $b=e^{i\beta} \in \partial \mathbb D$ and propagates along geodesic lines which are represented as circular segments on the Poincaré disk. The wave fronts perpendicular to the geodesic lines are also circular and are referred to in the literature as \textit{horocycles}~\cite{anderson2005hyperbolic}. Note that we do not impose any boundary conditions, that is, Eq.~(\ref{planewave}) is understood as an eigenfunction on the infinite lattice.

%---------------------------------------------------------------------
\subsection{Oscillatory and Exponential Eigenmodes}
%---------------------------------------------------------------------

In many applications, one is primarily interested in oscillating waves of the form (\ref{planewave}), where the momentum $k$ is real. This applies in particular to experimental realizations of electronic circuits~\cite{lee2018topolectrical,helbig2020generalized,dong2021topolectric,lenggenhager2021electric,lenggenhager2022simulating,chen2023hyperbolic,gluscevich2023dynamic}.

In the context of the AdS/CFT correspondence, however, where the Klein-Gordon equation is often studied as a fundamental toy model, the primary focus lies on \textit{exponentially decaying} eigenfunctions with imaginary momentum $k=\pm i \sqrt{1+M^2}$, since only then do the eigenmodes exhibit the required asymptotic behavior
\begin{equation}
\psi(z) \sim A(1-z \overline z)^{1-\Delta}+B(1-z \overline z)^{\Delta}
\end{equation}
for $|z|\to 1$. Here, $\Delta=\frac12(1+\sqrt{1+M^2})$ denotes the scaling dimension of the boundary operator that is holographically dual to $\psi$ \cite{ammon2015gauge}. Interestingly, these modes are known to remain stable even for negative squared masses above the Breitenlohner-Freedman bound~\cite{breitenlohner1982stability,basteiro23breitenloher}
\begin{equation}
M^2 \;\geq\; M_{\text{MB}}^2=-1\,.
\end{equation}
As our results presented below concern mainly the exponential modes, it will be convenient in the following to redefine the momentum $k$ by
\begin{equation}
\label{mudef}
\mu \;:=\;-\frac{1}{2}(1+ik)\,.
\end{equation}
Using this notation the continuum eigenfunction (\ref{planewave}) emitted from the source $b = e^{i \beta} \in \partial \mathbb D$ and dimension $\Delta=\mu+1$ simply reads
\begin{equation}
\label{ContinuumPlaneWave2}
\psi_{\mu,b}(z) = \left( \frac{|b-z|^2}{1-|z|^2} \right)^{\mu}
\end{equation}
and the corresponding eigenvalue is given by
\begin{equation}
\label{MuEigenvalue}
\lambda = -4\mu(\mu+1).
\end{equation}
%

%---------------------------------------------------------------------
\subsection{Radially Symmetric Eigenfunctions}
%---------------------------------------------------------------------

Since the eigenvalue $\lambda$ does not depend on the location $b = e^{i\beta}$ of the source on the boundary, the corresponding radial eigenfunctions can be obtained by a Fourier transformation in the angle $\beta$. Here one obtains
\begin{equation}
\begin{split}
u_k^m(z) &\;:=\;\frac{1}{2\pi} \int_0^{2\pi} \textrm{d}\beta \, e^{i m \beta}\psi_{k,e^{i \beta}} \\
&\;\propto\; e^{i m \phi} P^m_{-\frac12(1+i k)}\Bigl( \frac{1+r^2}{1-r^2} \Bigr)\,,
\end{split}
\end{equation}
where $m \in \mathbb N_0$ and $z=r e^{i\phi}$, while $P_\ell^m$ denotes the associated Legendre functions. An example of a radial eigenfunction is shown in the right panel of Fig.~\ref{fig:Eigensolutions}.

In the case of exponential modes, where we use the notation $\mu = -\frac{1}{2}(1+ik)$ defined in (\ref{mudef}), the radial eigenfunctions reads:
\begin{equation}
\label{ContinuumRadialWave}
u_{\mu}^m(z) \;\propto\; e^{i m \phi} P^m_{\mu}\Bigl( \frac{1+r^2}{1-r^2} \Bigr)\,.
\end{equation}
%
%=====================================================================
\section{Lattice Laplacian}
\label{sec:LatticeLaplacian}
%=====================================================================
%
Up to this point we have reviewed the eigenfunctions of the Laplace-Beltrami operator on the Poincaré disk in the continuum case. We now turn to a discretization of the hyperbolic space where we are going to study the analogue of the Laplace-Beltrami operator on the lattice.
%
%---------------------------------------------------------------------
\subsection{Definition of the Lattice Laplacian}
%---------------------------------------------------------------------
%
Before we define the hyperbolic lattice Laplacian, let us briefly recall the discrete Laplacian in the Euclidean case. On a one-dimensional line with equidistant vertices at positions $x_j=jh$, the finite-difference operator
\begin{equation}
\Delta^{[h]} f(x_j) = \frac{1}{h^2}\bigl( f(x_{j+1})+f(x_{j-1})-2f(x_j) \bigr)
\end{equation}
is known to approximate the second derivative $f''(x_j)$. More generally, on a regular hypercubic lattice with vertices at $\vec{x}_j~\in~\mathbb{R}^n$ and constant lattice spacing $h$, the discrete Laplacian $\Delta^{[h]}$ is defined as
\begin{equation}
\label{EuclideanLatticeLaplacian}
\Delta^{[h]} f(\vec{x}_j) = \frac{1}{h^2}\sum_k\left( A_{jk}-2n\delta_{jk} \right)f(\vec{x}_k)\,,
\end{equation}
where $A_{jk}$ denotes the adjacency matrix which is $1$ if the nodes $j$ and $k$ are connected by an edge and $0$ otherwise. The lattice Laplacian $\Delta^{[h]}$ is known to approximate the continuum Laplacian as $\Delta^{[h]} f(\vec{x}_j) = \nabla^2 f(\vec{x}_j) + \mathcal{O}(h^2)$.

\begin{figure}
\includegraphics[width=50mm]{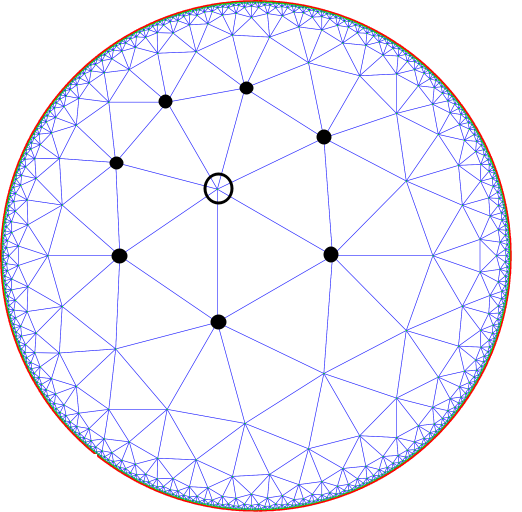}
\caption{$\{3,7\}$ tiling of the Poincaré disk. The vertices enumerated by $j$ are located at the points $z_j\in\mathbb D$. The lattice Laplacian applied to a function $f(z_j)$ living on the vertices is evaluated at a given vertex (hollow circle) and its nearest neighbors (solid dots) according to Eq.  (\ref{HyperbolicLatticeLaplacian}).
\label{fig:laplacian}}
\end{figure}

Now, let us turn to hyperbolic lattices on the two-dimensional Poincaré disk in the form of isometric $\{p,q\}$-tilings~\cite{coxeter1957crystal}. Such tilings consist of regular $p$-gons and vertices with coordination number $q$, satisfying the inequality $(p-2)(q-2)>4$, and can easily be generated using software packages such as~\cite{hypertiling}. An example of such a tiling is shown in Fig.~\ref{fig:laplacian}. Unlike Euclidean lattices, where the lattice spacing can be chosen freely, hyperbolic lattices have a fixed geodesic edge length.

Denoting the positions of the vertices on the Poincaré disk as $z_j\in\mathbb{D}$, the lattice Laplacian $\Delta^{\{p,q\}}$, which approximates the Laplace-Beltrami operator $\Delta_g$ in (\ref{LaplaceBeltrami}), is defined in analogy to (\ref{EuclideanLatticeLaplacian}) as
\begin{equation}
\label{HyperbolicLatticeLaplacian}
\Delta^{\{p,q\}} f(z_j) = \frac{1}{\mathcal{N}}\sum_k\left( A_{jk}-q\delta_{jk} \right)f(z_k)\,.
\end{equation}
Here $\mathcal N$ is a normalization factor given by~\cite{lenggenhager2022simulating}
\begin{equation}
\label{Normalization}
\mathcal{N} = \frac{1}{4}q h^2
\end{equation}
with
\begin{equation}
\label{RadiusH}
h \;=\; \left(1-\frac{\sin^2 (\pi/q)}{\cos^2 (\pi/p)}\right)^{1/2}\,.
\end{equation}
The objective of this work is to find eigenvectors $\Psi_{\mu,b}(z_j)$ of the eigenvalue problem
\begin{equation}
\label{LatticeEigenvalueProblem}
\Delta^{\{p,q\}} \Psi_{\mu,b}(z_j) = -\Lambda_\mu \, \Psi_{\mu,b}(z_j)\,,
\end{equation}
which are analogous to the plane-wave eigenfunctions (\ref{ContinuumPlaneWave2}) in the continuum case.
%
%---------------------------------------------------------------------
\subsection{Special Cases with Trivial Solutions}
%---------------------------------------------------------------------
%
In the Euclidean case, the continuum Laplacian $\Delta$ and its lattice counterpart $\Delta^{[h]}$ defined in Eq.~(\ref{EuclideanLatticeLaplacian}) are known to share the \textit{same} eigenfunctions
\begin{equation}
\psi_{\vec k}(\vec x)=e^{\pm i \vec k\cdot \vec x}
\end{equation}
independent of the lattice spacing $h$. The effect of the discretization is only reflected in a modification of the corresponding eigenvalues, namely
\begin{eqnarray}
\label{NormalDispersion}
\Delta \psi_{\vec k}(\vec x) &=& -k^2 \psi_{\vec k}(\vec x)\\
\label{DeformedDispersion}
\Delta_h \psi_{\vec k}(\vec x) &=& -\frac{4}{h^2}\Bigl( \sum_j \sin^2\frac{h k_j}{2}\Bigr) \psi_{\vec k}(\vec x)\,.
\end{eqnarray}
As can be seen, (\ref{DeformedDispersion}) tends to (\ref{NormalDispersion}) in the continuum limit $h \to 0$.

In the hyperbolic case, this coincidence of the continuum and lattice wave function is generally absent and the eigenfunctions on the lattice turn out to be highly nontrivial. Nonetheless, we have discovered specific exceptional values of $\mu$ where the coincidence still persists. More precisely, we have found that (\ref{LatticeEigenvalueProblem}) is solved by
\begin{equation}
\label{SpecialSolutions}
\Psi_{\mu,b}(z_j)=\psi_{\mu,b}(z_j) \qquad \text{for } \mu=0,1,\ldots,q-1
\end{equation}
with the eigenvalue
\begin{equation}
    \label{eq:GenZeroEigenvalues}
    \Lambda_\mu^{[0]}=\frac{q}{\mathcal N}\left( 1-P_\mu \Bigl( \tfrac{1+h^2}{1-h^2} \Bigr)\right)\,.
\end{equation}
This important result will be explained and proven in the following sections.

If we formally take $h \to 0$, the eigenvalue (\ref{eq:GenZeroEigenvalues}) tends to $-4\mu(\mu+1)$ in agreement with Eq.~(\ref{MuEigenvalue}). This justifies the choice of the normalization $\mathcal N$ in Eq.~(\ref{Normalization})~\cite{lenggenhager2022simulating}. However, one should keep in mind that on a hyperbolic lattice the radius $h$ has a certain constant value given in (\ref{RadiusH}) that cannot be taken to zero. Therefore, there seems to be an element of arbitrariness in the definition of the normalization which to our knowledge is not yet fully understood.

%=====================================================================
\section{Concept of Local Inclination}
\label{sec:Inclination}
%=====================================================================

An extensive numerical study of the eigenvalue problem for general positive and non-integer values of $\mu$ revealed that the lattice eigenfunctions that are analogous to plane waves differ only slightly from the continuum eigenfunctions~(\ref{ContinuumPlaneWave2}). In particular they seem to exhibit the same type of exponential characteristics on large scales. This suggests that the lattice eigenfunctions $\Psi_{\mu,b}(z_j)$ differ from the continuum eigenfunctions $\psi_{\mu,b}(z_j)$ only by some \textit{correction factor} $\chi_j$, which is expected to be close to $1$ everywhere on the infinite lattice:
\begin{equation}
\label{CorrectionFunction}
\Psi_{\mu,b}(z_j) \;=\; \psi_{\mu,b}(z_j)\, \chi_j\,.
\end{equation}
So far, this ansatz does not impose any restrictions. However, as a central hypothesis of this paper, we propose that the correction factors $\chi_j$ exhibit a very specific dependence on the local configuration of the lattice. This leads us to the \textit{inclination hypothesis}, as will be explained in the following.
%
%---------------------------------------------------------------------
\subsection{Inclination Hypothesis}
%---------------------------------------------------------------------
%
%
\begin{figure}
\includegraphics[width=85mm]{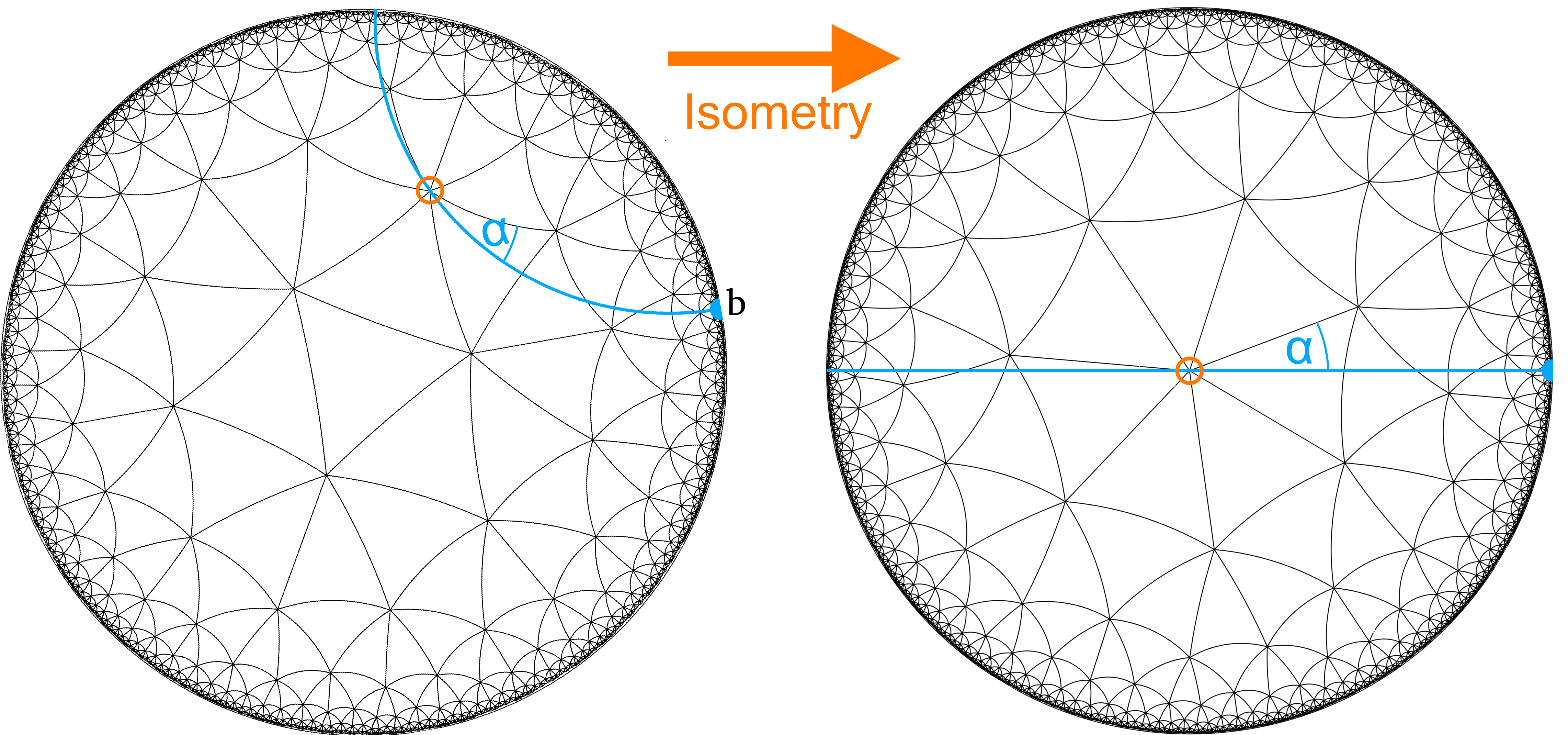}
\caption{
\label{fig:inclination}
Left: $\{3,7\}$-tiling with a geodesic line (blue) originating at $b\in \partial \mathbb D$ which crosses the vertex marked by the hollow circle. $\alpha$ is defined as the angle between the geodesic line and the next edge in counterclockwise direction. Right: Mapping the vertex conformally to 0 and the source to 1 (see text).
}
\end{figure}

The numerical observation that the correction factors $\chi_j$ are close to $1$ on the entire lattice suggests that the asymptotic behavior of the eigenfunction $\Psi_{\mu,b}(z_j)$ in (\ref{CorrectionFunction}) is already correctly captured by the continuous eigenfunction $\psi_{\mu,b}(z_j)$ while the factors $\chi_j$ can be understood as corrections caused by the local configuration of the lattice around the vertex $z_j$. However, since all vertices are isometrically equivalent, the only feature in which they can differ is the direction in which they 'see' the source~$b$ from which the plane wave emanates, relative to the vertex itself and its nearest neighbors.

Therefore, we conjecture that the correction factors $\chi_j$ can only depend on the angle $\alpha_j$ at which a geodesic line emanating from the source $b$ intersects the vertex in relation to its nearest neighbors. As demonstrated in Fig.~\ref{fig:inclination}, this angle can be determined as follows. First, an arbitrary vertex $z_j$ is selected, marked by the hollow circle in the figure. Then the source $b$ and $z_j$ are connected by a geodesic line, represented as a circular segment on the Poincaré disk. The angle $\alpha_j$ is then the intersection angle of this geodesic line with the first outgoing edge of the vertex in the counterclockwise direction.

Since both the geodesic line and the edges exhibit a curvature in their representation on the Poincaré disk, it is advantageous to apply an isometry to bring $b$ to $1$ and the vertex to the center of the disk, whereby both lines become straight. Since the conformal isometry preserves angles, this allows us to easily read off the angle $\alpha_j$, as shown in the right panel of Fig.~\ref{fig:inclination}.

Since $\alpha_j \in [0,2\pi/q)$, it will be convenient in the following to rescale this angle, defining the \textit{inclination}
\begin{equation}
\label{Inclination}
\tau_j \;:= \; \frac{q}{2\pi}\alpha_j \;\in\; [0,1)\,.
\end{equation}
For a given position~$b$ of the source, one can calculate the local inclination $\tau_j$ at every vertex~$j$, as will be explained in the following.
%
%---------------------------------------------------------------------
\subsection{Explicit Calculation of the Inclination}
\begin{figure}
\centering\includegraphics[width=85mm]{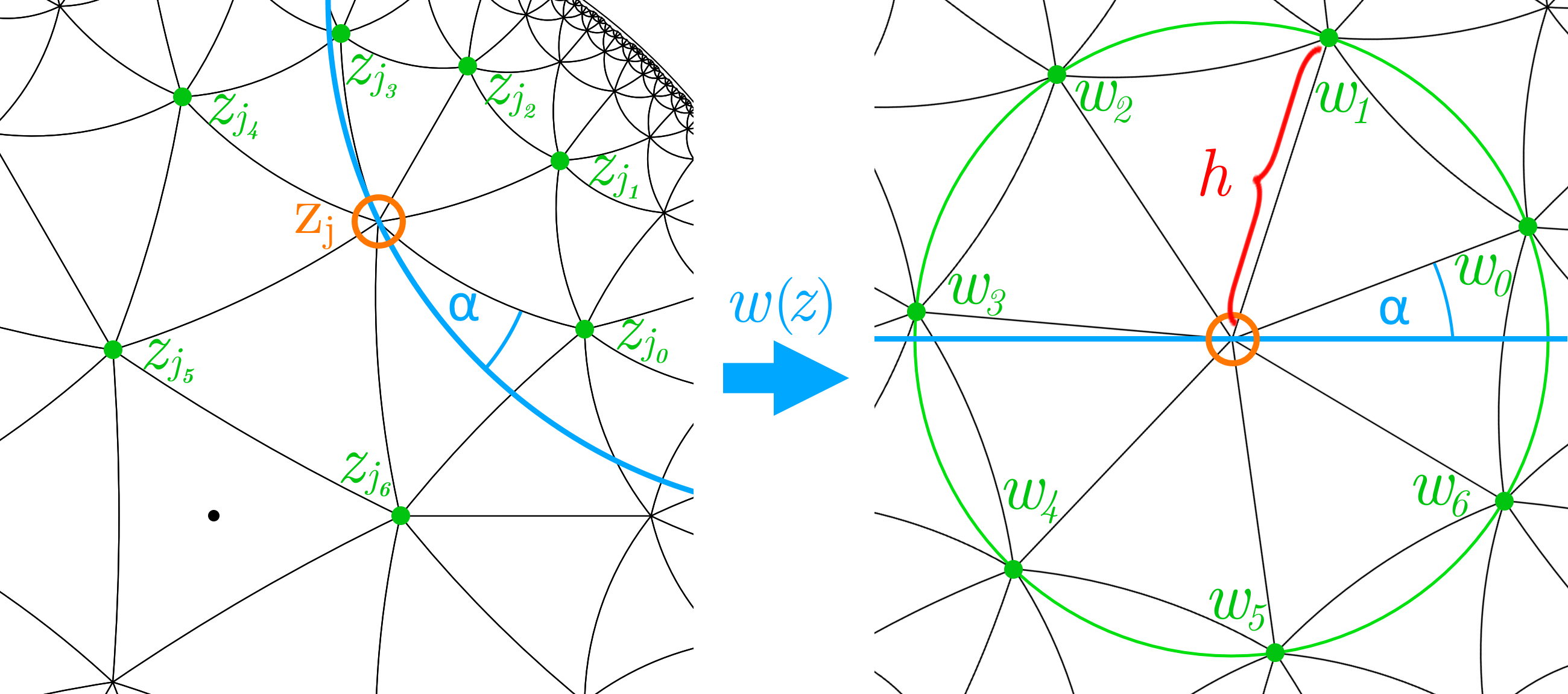}
\caption{
\label{fig:taucalc}
Mapping the cell with the vertex $j$ to the center in the example of a $\{3,7\}$-lattice.
}
\end{figure}

%---------------------------------------------------------------------
%
Let us choose an arbitrary vertex $j$ on the lattice with the coordinate $z_j\in\mathbb D$ and let us denote by $z_{j_0},z_{j_1},\ldots,z_{j_{q-1}}$  the coordinates of the neighboring vertices $j_0,j_1,\ldots,j_{q-1}$, as sketched in Fig.~\ref{fig:taucalc}. Furthermore, let $b=e^{i\beta}\in\partial\mathbb D$ be the coordinate of the source of the plane wave at the boundary. As outlined above, let us apply the isometry
\begin{equation}
\label{SpecialIsometry}
z \mapsto w(z)=\frac{(1-b\overline z_j)(z-z_j)}{(b-z_j)(1-z\overline z_j)}\,,
\end{equation}
which maps $b$ to $1$ and $z_j$ to the disk's center at $0$. As shown in the figure, the new positions of the nearest neighbors $w_n=w(z_{j_n})$ are then uniformly distributed along a circle
\begin{equation}
\label{DefWn}
w_n  \;=\; h \, e^{i(\alpha_j + 2\pi n/q)} \qquad (n=0,\ldots,q-1)
\end{equation}
with the radius $h$ given in (\ref{RadiusH}). As can be seen in the right panel, the $q$ edges are now straight and form a regular star, which is tilted against the horizontal geodesic line by the angle $\alpha_j$. Choosing a nearest neighbor with the index  $n=0,\ldots,q-1$, the angle $\alpha_j$ is then given by
\begin{equation}
e^{i q \alpha_j}\;=\;\Bigl(\frac{w_n}{h}  \Bigr)^q\,,
\end{equation}
As can be seen, the actual choice of $n$ is irrelevant since it drops out when taking the $q^\textrm{th}$ power. Therefore, we can select just one arbitrary nearest neighbor $j'=j_n$ of the vertex $j$ and denote its coordinate as $z'_j$. Since we have defined the inclination as $\tau_j = \frac{q}{2\pi}\alpha_j$, we arrive at
\begin{equation}
\label{inclinationRelation}
e^{2\pi i \tau_j} \;=\; \Bigl[ \frac 1 h \, \frac{(1-b \overline z_j)(z'_j-z_j)}{(b-z_j)(1-z'_j\overline z_j)} \Bigr]^q\,,
\end{equation}
where the square bracket encapsulates a complex number on the unit circle. Taking the argument allows us to disregard the factor $\frac{1}{h}$, leading to the final expression for the local inclination:
\begin{equation}
\label{inclinationResult}
\tau_j \;=\; \frac{1}{2\pi} \arg \Bigl(\Bigl[ \frac{(1-b \overline z_j)(z'_j-z_j)}{(b-z_j)(1-z'_j\overline z_j)} \Bigr]^q\Bigr)\,.
\end{equation}
Here $\tau_j$ is evaluated modulo 1 in the range $[0,1)$. In practice, once we have generated a hyperbolic tiling with vertex coordinates $z_j$ and defined the location of the source $b \in \partial \mathbb D$, this formula allows us to compute the inclination parameter $\tau_j$ at every vertex $j$ of the infinite lattice simply by choosing just one of its nearest neighbors $j'$ at the position $z_j'$ and evaluating~(\ref{inclinationResult}).
%
%=====================================================================
\section{Calculation of the Correction Function $\chi(\tau)$}
\label{sec:CorrectionFunction}
%=====================================================================
%
As the main hypothesis of this work, we suggest that the correction factors $\chi_j$ depend only on the local inclination $\tau_j$, that is, we suggest that there exists a \textit{correction function} $\chi(\tau): [0,1)\to\mathbb C$ such that
\begin{equation}
\label{MainAssumption}
\chi_j = \chi(\tau_j).
\end{equation}
If this assumption turns out to be correct and if we succeed to calculate the function $\chi(\tau)$ explicitly, this will give us the exact eigenfunction on the entire lattice.
%
%---------------------------------------------------------------------
\subsection{Finite Difference Equation for $\chi(\tau)$}
%---------------------------------------------------------------------
%
We now present a linear difference equation which determines the correction function $\chi(\tau)$. Starting point is the eigenvalue problem~(\ref{LatticeEigenvalueProblem})
\begin{equation}
\label{EigenvalueProblem}
\Delta^{\{p,q\}}\Psi_{\mu,b}(z_j) \;=\; -\Lambda_\mu\Psi_{\mu,b}(z_j)\,,
\end{equation}
where $\Delta^{\{p,q\}}$ denotes the hyperbolic lattice Laplacian defined in (\ref{HyperbolicLatticeLaplacian}). Using the abbreviation $\Psi_j:=\Psi_{\mu,b}(z_j)$, this eigenvalue problem can be recast as
\begin{equation}
\sum_k A_{jk}\Psi_k \;=\; \bigl(q-\mathcal N \Lambda_\mu\bigr) \Psi_j\,.
\end{equation}
where $A_{jk}$ is again the adjacency matrix and $\mathcal N$ is the normalization factor defined in (\ref{Normalization})-(\ref{RadiusH}).

As explained above, we assume that the lattice eigenfunction $\Psi_j$ can be expressed as
\begin{equation}
\Psi_j\;=\;\psi_j\,\chi(\tau_j)\,,
\end{equation}
where
\begin{equation}
\label{ContinuumEigenfunction}
\psi_j :=\psi_{\mu,b}(z_j) = \left( \frac{|b-z_j|^2}{1-|z_j|^2} \right)^{\mu}
\end{equation}
is the continuum eigenfunction and $\chi(\tau)$ is the correction function we aim to determine. As shown in Appendix~\ref{AppendixDifferenceEquation}, this ansatz leads directly to a linear difference equation of the form
\begin{equation}
\label{DifferenceEquation}
\sum_{n=0}^{q-1} R_n(\tau) \, \chi\bigl(\sigma_{n}(\tau)\bigr) \;=\; \bigl(q-\mathcal N \Lambda_\mu\bigr) \chi(\tau)\,,
\end{equation}
which is expected to hold for all inclinations $\tau\in[0,1)$. Here the function $R_n(\tau)$ is defined as
\begin{equation}
R_n(\tau) \;=\;
\biggl( \frac{\bigl|1-h Z^{\tau+n}\bigr|^2}{1-h^2} \biggr)^{\mu}
\end{equation}
with $Z=e^{2\pi i/q}$ while
\begin{equation}
\sigma_n(\tau) \;=\; \frac{1}{2\pi} \arg \left[ \Bigl( \frac{h-Z^{\tau+n}}{1-h Z^{\tau+n}} \Bigr)^q \right] \in [0,1)\,
\end{equation}
denotes the inclinations of the nearest neighbors indexed by $n=0,1,\ldots,q-1$. Eq.~(\ref{DifferenceEquation}) provides a linear relation between the function $\chi$ evaluated at the inclination $\tau$ and the same function evaluated at the $q$ different inclinations $\sigma_0(\tau),\ldots,\sigma_{q-1}(\tau)$ of the nearest neighbors. Thus, it can be considered as a non-local linear difference equation for the function $\chi(\tau)$.
%
%---------------------------------------------------------------------
\subsection{Numerical Iteration Scheme}
%---------------------------------------------------------------------
%
The difference equation (\ref{DifferenceEquation}) can be used to approximate the correction function $\chi(\tau)$ numerically. As detailed in Appendix~\ref{AppendixNumericalScheme}, if we discretize the correction function and if we start without correction by setting $\chi(\tau)=1$, we can use the difference equation to iterate $\chi(\tau)$ in such a way that it converges towards a stable solution.

\begin{figure}
\includegraphics[width=85mm]{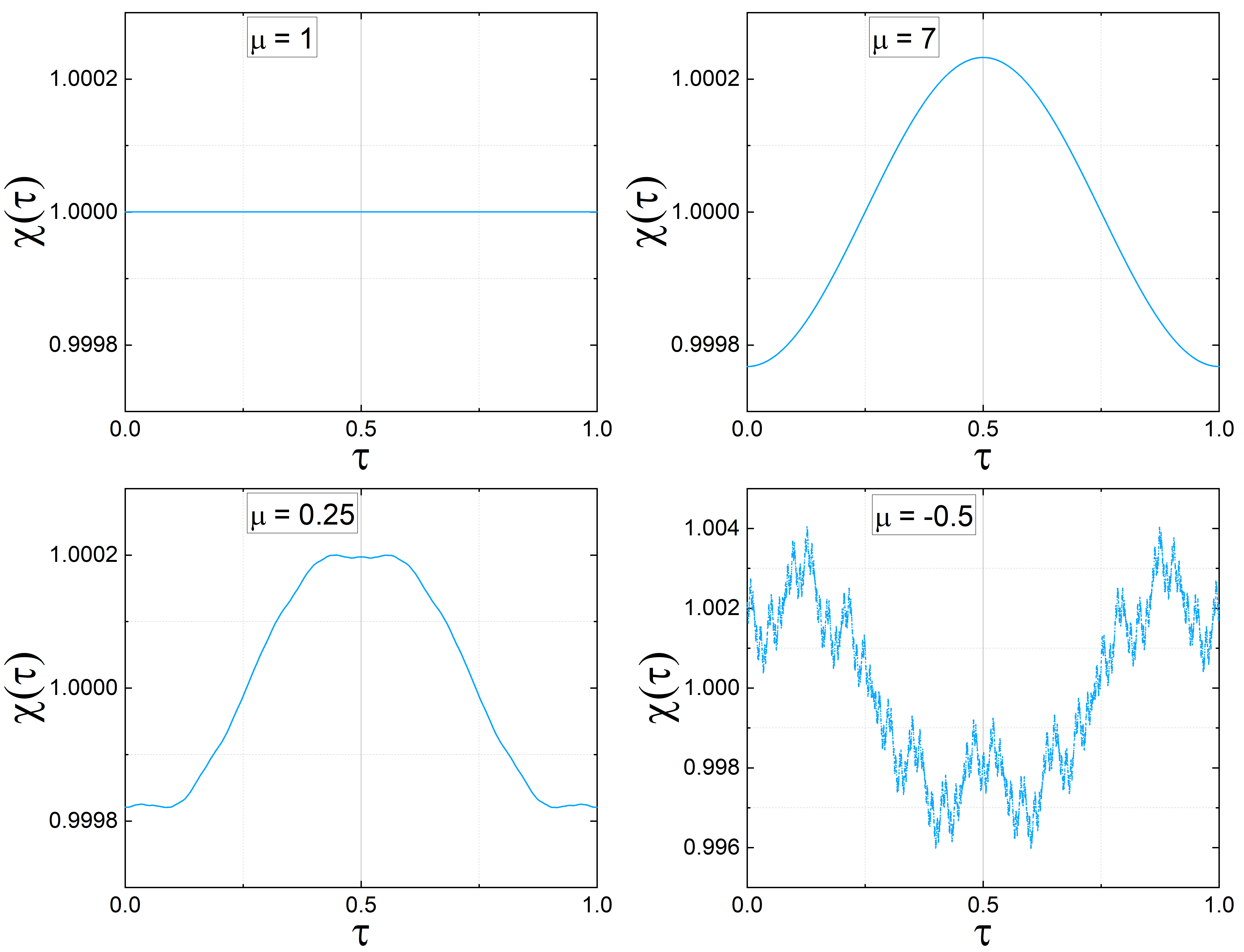}
\caption{
\label{fig:chi}
Numerically determined correction function $\chi(\tau)$ on a $\{3,7\}$-lattice for various values of $\mu$ (see text).
}
\end{figure}

Fig.~\ref{fig:chi} shows four examples of these numerical solutions. Our results confirm that for $\mu=0,1,\ldots,q-1$ no correction is needed, i.e. $\chi(\tau)=1$, in agreement with our earlier observation in Eq.~(\ref{SpecialSolutions}). In all other cases $\chi(\tau)$ differs only slightly from 1. For integer values $\mu=q,q+1,\ldots,2q-1$ we obtain a solution of the form $\chi(\tau)=1-A \cos(2\pi\tau)$. For larger integer values of $\mu$ more and more Fourier modes are involved. Generally, we observe numerically that for positive integer $\mu$ the function $\chi(\tau)$ involves $\lfloor \mu/q \rfloor$ Fourier modes.

For positive but non-integer values $\mu\in\mathbb R$ the solutions are stable but they seem to involve infinitely many Fourier modes. Nevertheless, as the lower left panel in Fig.~\ref{fig:chi} suggests, for $\mu=1/4$ the function $\chi(\tau)$ is still well-defined.

For negative values of $\mu$, however, the situation is different. Here the iteration scheme still converges, but the resulting function $\chi(\tau)$ turns out to be highly nontrivial. As can be seen in the lower right panel of the figure, the irregular form reminds one of the Weierstrass function, which is known in mathematics as an example of a continuous but nowhere differentiable function.

\begin{figure}
\includegraphics[width=75mm]{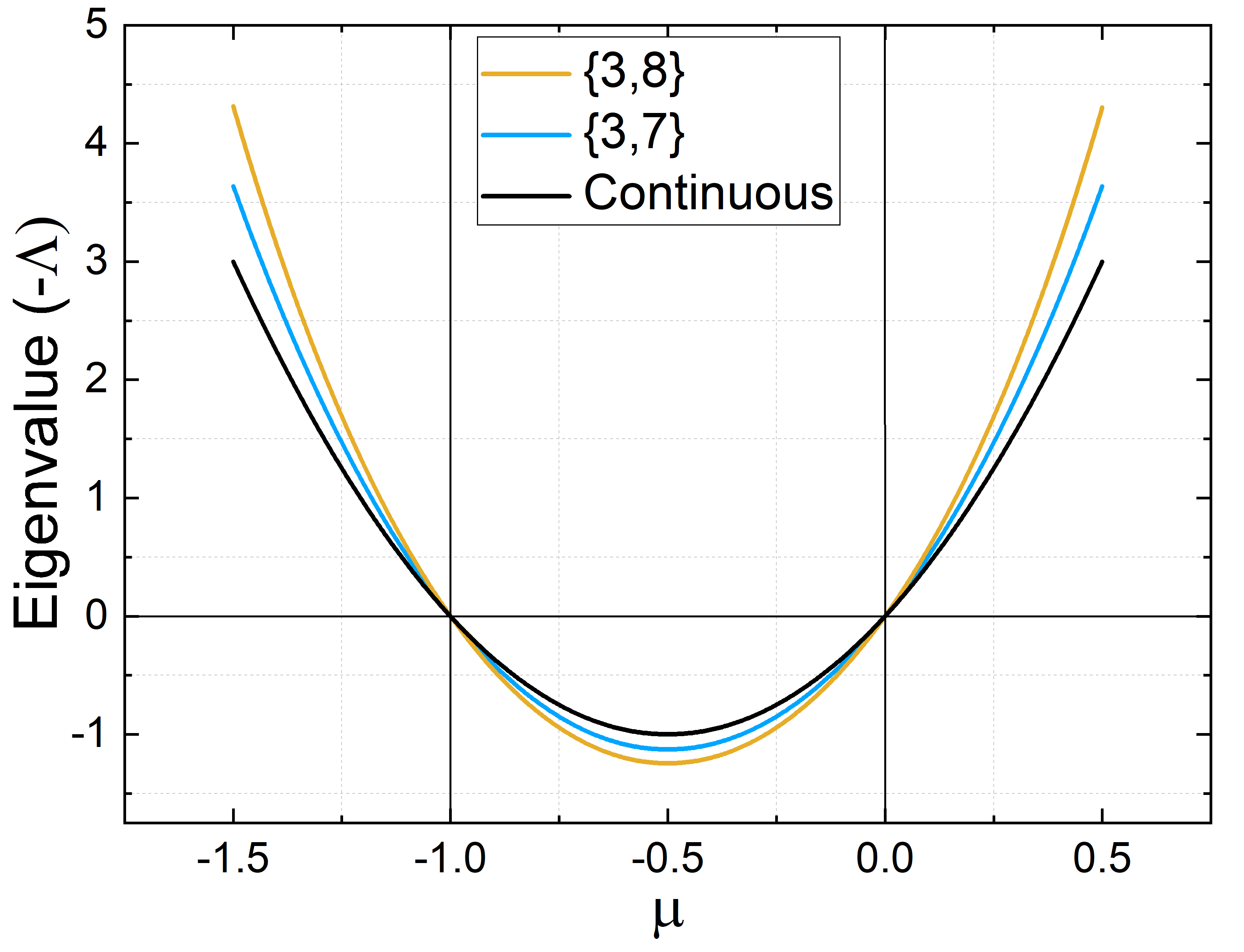}
\caption{
Numerically determined eigenvalue $-\Lambda_\mu$ as a function of $\mu$ on two different lattices and in the continuum case.
\label{fig:eigenvalueplot}}
\end{figure}

Fig.~\ref{fig:eigenvalueplot} shows the numerically determined eigenvalue $(-\Lambda)$ as a function of $\mu$ on two different lattices in comparison with the continuum case. As can be seen, the eigenvalue vanishes at $\mu=0$ and $\mu=-1$ in all cases. Moreover, we observe on both lattices the important symmetry $\Lambda_\mu=\Lambda_{-1-\mu}$, which will be proven in the following Section. Some explicit numerical eigenvalues are listed in Table~\ref{tab:eigenvalues}.

\begin{table}
\begin{center}
\def\arraystretch{1.3}
\begin{tabular}{c||c|c|c}
 $\mu$ & $\{3,7\}$ & $\{3,8\}$ & $\{4,8\}$ \\ %[0.5ex]
 \hline\hline
 $-0.5$ & $1.127627$ & $1.245405$ & $1.548876$ \\
 \hline
 $-0.25$ & $0.849605$ & $0.942441$ & $1.194023$ \\
 \hline
 $0.25$ & $-1.468768$ & $-1.686357$ & $-2.402387$ \\
 \hline
 $0.5$ & $-3.639259$ & $-4.304511$ & $-6.726293$ \\
 \hline
 $1$ & $-10.62388$ & $-13.65685$ & $-27.31370$ \\
 \hline
 $7$ & $-7.465861\times10^3$ & $-9.207830\times10^4$ & $-3.302949\times10^7$ \\
 \hline
 $8$ & $-2.083553\times10^4$ & $-3.980441\times10^5$ & $-3.583152\times10^8$ \\
 \hline \hline
 $h$            & $0.4969704$ & $0.6435943$ & $0.8408964$ \\
 \hline
 $\mathcal N$   & $0.4322143$ & $0.8284271$ & $1.4142136$
\end{tabular}
\end{center}
\caption{Numerically determined eigenvalues $\Lambda_\mu$ as well as the parameters $h$ and $\mathcal N$ for three different $\{p,q\}$-lattices. The numerical precision is $\pm 1$ in the last digit.
\label{tab:eigenvalues}
}
\end{table}

%=====================================================================
\section{Fourier Analysis}
\label{sec:FourierAnalysis}
%=====================================================================

%---------------------------------------------------------------------
\subsection{Fourier Expansion of the Correction Function}
%---------------------------------------------------------------------
%
The numerical observation that for integer $\mu$ the function $\chi(\tau)$ involves only a finite number of Fourier modes suggests to expand it as a Fourier series
\begin{equation}
\label{FourierAnsatz}
\chi(\tau) \;=\; \sum_{k=-\infty}^{+\infty} \gamma_k \; e^{2\pi i \tau k}\,,
\end{equation}
where $\gamma_k\in\mathbb C$ are Fourier coefficients with $\gamma_0:=1$. As shown in Appendix~\ref{AppendixFourierExpansion}, this converts the difference equation (\ref{DifferenceEquation}) into an eigenvalue problem for the Fourier coefficients:
\begin{equation}
\label{GammaEVProblem}
\sum_{k=-\infty}^{\infty} B_{j,k}\,\gamma_k \;=\; (q-\mathcal N \Lambda_\mu) \gamma_j\,.
\end{equation}
Here, $B$ is an infinite-dimensional matrix defined by
\begin{equation}
B_{j,k} = \frac{q \,(-1)^{qj}\,h^{q|j-k|}}{(1-h^2)^\mu}
\begin{cases}
\binom{\mu-qk}{\mu-qj}F_{qj,-qk} & \text{ if } j\geq k \\
\binom{\mu+qk}{\mu+qj}F_{-qj,qk} & \text{ if } j< k
\end{cases}
\end{equation}
where we used the abbreviation
\begin{equation}
\label{Abbreviation}
F_{a,b}={}_2F_1\bigl(a-\mu,\,b-\mu;\,1+a+b;\,h^2\bigr)
\end{equation}
and where $\binom{a}{b}=\tfrac{\Gamma(1+a)}{\Gamma(1+b)\Gamma(1+a-b)}$. Note that the matrix $B$ is transposed under the substitution $\mu \to -1-\mu$, proving the previously observed symmetry of the eigenvalues:
\begin{equation}
\Lambda_\mu =\Lambda_{-1-\mu}\,.
\end{equation}
It is remarkable that this symmetry is valid not only in the continuum but on any $\{p,q\}$-lattice.
%---------------------------------------------------------------------
\subsection{Exact Solution for $\mu\in\mathbb N$}
%---------------------------------------------------------------------
%
\begin{figure}
\includegraphics[width=55mm]{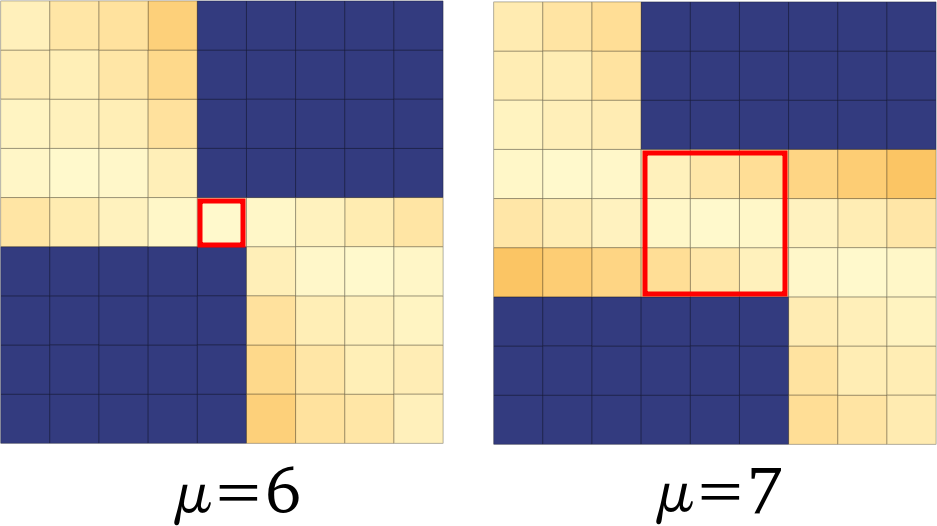}
\caption{Block structure of the matrix $B$ for $\mu=6$ and $\mu=7$ on a $\{3,7\}$-lattice. The panels show the central section of the matrix $B_{j,k}$ where $j,k\in\{-4,\ldots,4\}$. The dark fields indicate entries where the matrix is exactly zero. The blocks in the center marked in red color decouple in the corresponding eigenvalue problem.
\label{fig:blockmatrix}}
\end{figure}
Interestingly, for $\mu \in \mathbb N$ the matrix $B$ possesses a block structure which allows us to find exact solutions. More specifically, for given $\mu \in \mathbb N$, choosing $s = \lfloor \mu/q \rfloor \in \mathbb N$ such that $qs \leq \mu < q(s+1)$, we find that
\begin{equation}
\begin{cases}
B_{j,k}=0 & \text{ if } j>s \text{ and } k\leq s \\
B_{j,k}=0 & \text{ if } j<-s \text{ and } k\geq -s \\
B_{j,k}\neq0 &  \text{ otherwise}
\end{cases}
\end{equation}
This gives rise to a block structure of the matrix. This block structure is visualized in Fig.~\ref{fig:blockmatrix}, where the central part of the matrix is shown for $\mu=6$ and $\mu=7$ on a $\{3,7\}$-lattice. Each square in the figure stands for a matrix element. The dark blue squares indicate that the respective matrix elements are zero.

As can be seen, the central block matrix with indices in the range $|j|\leq s$ and $|k| \leq s$ (enclosed by a red square in the figure) decouples in the corresponding eigenvalue problem. This enables us to solve the corresponding eigenvalue problem separately within this block and to set all remaining Fourier coefficients $\gamma_k$ with $|k|>s$ to zero. Therefore, the resulting Fourier series (\ref{FourierAnsatz}) contains only $2s+1$ terms. This observation is one of the main results of this work.
\vspace{2mm}

\begin{table}
\begin{center}
\def\arraystretch{1.3}
\begin{tabular}{|c|c||c|c|}
\hline
$\{p,q\}$ & $\;\mu\;$  & $\Lambda_\mu$ & $\left\{ \gamma_0,\gamma_1 \right\}$\\
\hline
$\{3,7\}$ & $1$ & $-10.6239$ & $\left\{1,0\right\}$ \\
& $2$ & $-42.3251$ & $\left\{1,0\right\}$ \\
& $6$ & $-2.69424\times 10^3$ & $\left\{1,0\right\}$ \\
& $7$ & $-7.46586\times 10^3$ & $\left\{1,-1.16212\times 10^{-4}\right\}$ \\
& $8$ & $-2.08355\times 10^4$ & $\left\{1,-5.41973\times 10^{-4}\right\}$ \\
& $9$ & $-5.8503\times 10^4$ & $\left\{1,-1.46462\times 10^{-3}\right\}$ \\
\hline
$\{3,8\}$ & $1$ & $-13.6569$ & $\left\{1,0\right\}$ \\
& $2$ & $-69.9411$ & $\left\{1,0\right\}$ \\
& $6$ & $-2.15029\times 10^4$ & $\left\{1,0\right\}$  \\
& $7$ & $-9.20783\times 10^4$ & $\left\{1,0\right\}$  \\
& $8$ & $-3.98044\times 10^5$ & $\left\{1,5.15986\times 10^{-5}\right\}$ \\
& $9$ & $-1.73383\times 10^6$ & $\left\{1,2.58365\times 10^{-4}\right\}$ \\
\hline
$\{4,8\}$ & $1$ & $-27.3137$ & $\left\{1,0\right\}$ \\
& $2$ & $-279.765$ & $\left\{1,0\right\}$ \\
& $6$ & $-3.07439\times 10^6$ & $\left\{1,0\right\}$  \\
& $7$ & $-3.30295\times 10^7$ & $\left\{1,0\right\}$  \\
& $8$ & $-3.58315\times 10^8$ & $\left\{1,7.28970\times 10^{-5}\right\}$ \\
& $9$ & $-3.91747\times 10^9$ & $\left\{1,3.50096\times 10^{-4}\right\}$ \\
\hline
\end{tabular}
\end{center}
\caption{Selected numerical results for the eigenvalue $\Lambda$ and the Fourier coefficients $\{\gamma_k\}$ for integer $\mu$ on different lattices.
\label{tab:xxx}
}
\end{table}
The block structure enables us to prove various special cases exactly. For instance, for $\mu=0,1,\ldots q-1$, we can now easily prove the special case discovered in Eqs. (\ref{SpecialSolutions})-(\ref{eq:GenZeroEigenvalues}), where the lattice and continuum eigenfunctions coincide. In this case we have $s=0$, obtaining a one-dimensional block matrix. This block matrix has the trivial eigenvector $\gamma_0=1$ and the eigenvalue is given by the matrix element itself:
\begin{equation}
 \Lambda^{[s=0]}_\mu =  \frac{q}{\mathcal N}\Bigl( 1-\frac{1}{(1-h^2)^\mu} {}_2F_1(-\mu,-\mu;\, 1;\,h^2) \Bigr)\,.
\end{equation}
For $\mu=q,\ldots,2q-1$, where $s=1$, we find an explicit exact solution for the lattice eigenfunction, which can be found in Appendix~\ref{AppendixExactSolution}.

Eq.~(\ref{GammaEVProblem}) is also beneficial for non-integer $\mu>0$, as we can approximate the solution numerically by truncating the matrix $B$ at a large but finite size and diagonalize it numerically. This gives a very good approximation of a finite number of Fourier coefficients $\gamma_k$, allowing us to compute a reliable approximation of the eigenmodes on the infinite lattice.

%---------------------------------------------------------------------
\subsection{Radial Eigenfunctions on the Lattice}
%---------------------------------------------------------------------

Having calculated the plane-wave lattice eigenfunctions originating from $b = e^{i\beta} \in \partial \mathbb D$, the corresponding radially symmetric solutions analogous to $u_{\mu}^m(z)$ in Eq. (\ref{ContinuumRadialWave}) can be obtained through a simple Fourier transformation in the angle $\beta$:
\begin{equation}
U_{\mu}^m(z_j)\;=\;\frac{1}{2\pi}\int_0^{2\pi} e^{i m \beta}\,\Psi_{\mu,e^{i\beta}}(z_j)
\,.
\end{equation}
In this context, a straightforward calculation (not shown here) leads to the following explicit expression in terms of the Fourier coefficients $\gamma_k$:
\begin{equation}
\begin{split}
U_{\mu}^m(z_j) &=\frac{1}{(1-|z_j|^2)^\mu}\,
\sum_{k=-\infty}^\infty \gamma_k
\left[ \frac{z_j-z'_j}{h (1-z'_j \bar z_j)} \right]^{qk} \times \\
&
\times
% \binom{\mu-q k}{\mu-m}\,{}_2 F_1\bigl(-\mu-qk,\;-\mu+m;\;1+m-qk;\;|z_j|^2\bigr)
\begin{cases}
z_j^{m-q k}\binom{\mu-q k}{m-q k}\, F_{m,-qk}(|z_j|^2) & \text{ if } m \geq q k \\
\overline z_j^{q k-m}\binom{\mu+q k}{q k-m}\,  F_{-m,qk}(|z_j|^2)  & \text{ if } m < q k \\
\end{cases}
\end{split}
\end{equation}
where $F_{a,b}(\xi)={}_2F_1\bigl(a-\mu,\,b-\mu;\,1+a+b;\,\xi\bigr)$.

%=====================================================================
\section{Conclusions}
\label{sec:Conclusions}
%=====================================================================

%
In the present paper, we have introduced a method for computing specific eigenfunctions of the hyperbolic lattice Laplacian on the Poincaré disk. These eigenfunctions are analogous to plane waves in the continuum case. Our approach hinges on the assumption that such eigenfunctions deviate from the continuum solutions by a correction function $\chi(\tau)\approx 1$ which depends solely on what we call the local inclination $\tau$ of the respective vertex relative to the propagation direction of the wave.

We have provided a numerical iteration scheme alongside an analytical method that allows us to determine the correction function $\chi(\tau)$ for given lattice parameters $\{p,q\}$ and the exponent $\mu$. Furthermore, we have demonstrated that for $\mu\in\mathbb N$, this correction function incorporates only finitely many Fourier modes, which allows various special cases to be solved exactly.

It is noteworthy that our approach works for any $\{p,q\}$-tiling and does not rely on edge pairing and the existence of a Fuchsian translation subgroup. Moreover, once the correction function $\chi(\tau)$ is determined on the interval [0,1), it promptly allows us to calculate the eigenfunction on the entire infinite lattice without cutoff.

Despite these advances, many questions remain open. In the current study, we focused on (unnormalizable) exponential modes with $\mu>0$, corresponding to imaginary momenta $k=i(1+2\mu)$. As the numerical results suggest, for $\mu<0$ the form of the eigenfunctions is significantly more irregular, and the interpretation of these findings is not yet clear. It would also be intriguing to study oscillatory modes with real momenta. Preliminary tests suggest that the proposed methods remain partially functional, but this requires further detailed exploration. Additionally, we analyzed only the largest eigenvalue of the difference function which leads to a plane-wave analogue, so far neglecting many other possible solutions. Finally, the correlation between our results and hyperbolic band theory and Fuchsian symmetry is still to be clarified.

%=====================================================================
\begin{acknowledgments}
We are grateful to J.~Erdmenger for fruitful discussions.
\end{acknowledgments}
%=====================================================================

%=====================================================================
\appendix
%=====================================================================

%=====================================================================
\section{Linear Difference Equation for the Correction Function}
\label{AppendixDifferenceEquation}
%=====================================================================
%
In this Appendix we present details for deriving the linear difference equation (\ref{DifferenceEquation}) for the correction function~$\chi(\tau)$. Starting point is the eigenvalue problem~(\ref{LatticeEigenvalueProblem})
\begin{equation}
\sum_k A_{jk}\Psi_k \;=\; \bigl(q-\mathcal N \Lambda_\mu\bigr) \Psi_j\,,
\end{equation}
where $\Psi_j:=\Psi_{\mu,b}(z_j)$ is the lattice eigenmode, $A_{jk}$ denotes the adjacency matrix, and $\mathcal N = \frac14 q h^2$ is the normalization factor. As explained above, we rewrite the lattice eigenmode $\Psi_j$ as
\begin{equation}
\Psi_j=\psi_j\chi_j\,,
\end{equation}
where
\begin{equation}
\label{ContinuumEigenfunctionAppendix}
\psi_j :=\psi_{\mu,b}(z_j) = \left( \frac{|b-z_j|^2}{1-|z_j|^2} \right)^{\mu}
\end{equation}
is the continuum eigenfunction and where the $\chi_j$ are correction factors we aim to determine. This ansatz turns the eigenvalue problem into
\begin{equation}
\sum_k A_{jk}\psi_k\chi_k = (q-\mathcal N \Lambda_\mu) \psi_j\chi_j\,.
\end{equation}
Since $\psi_j\neq 0$, we can divide by $\psi_j$, yielding an equivalent eigenvalue problem for the corrections $\chi_j$:
\begin{equation}
\label{EigenvalueProblem1a}
\sum_k A_{jk}\frac{\psi_k}{\psi_j}\chi_k \;=\;
\bigl(q-\mathcal N \Lambda_\mu\bigr) \chi_j\,.
\end{equation}
Since the adjacency matrix runs over $q$ nearest neighbors, we can rewrite this eigenvalue problem as
\begin{equation}
\label{EigenvalueProblem1b}
\sum_{n=0}^{q-1} R_{j,n}\,\chi_{j_n} \;=\;
\bigl(q-\mathcal N \Lambda_\mu\bigr) \chi_j\,,
\end{equation}
where $j_n$ denotes the index of the $n^\text{th}$ nearest neighbor of the vertex $j$ and where
we defined the ratio of the continuum eigenmodes
\begin{equation}
R_{j,n} \;:=\; \frac{\psi_{j_{n}}}{\psi_{j}}\,.
\end{equation}
It can be shown that this ratio is in fact an isometric invariant, i.e., it remains unaltered under any isometry of the form $z \mapsto w(z)=e^{i\eta}\tfrac{a-z}{1-z \overline a}$:
\begin{equation}
R_{j,n} \;=\;\frac{\psi_{j_n}}{\psi_j}\;=\;\frac{\psi_{\mu,b}(z_{j_n})}{\psi_{\mu,b}(z_j)}\;=\; \frac{\psi_{\mu,w(b)}(w(z_{j_n}))}{\psi_{\mu,w(b)}(w(z_j))}\,.
\end{equation}
We apply this invariance in the case of the isometry (\ref{SpecialIsometry}) in order to move the source $b \in \partial \mathbb D$ to $1$ and the vertex at $z_j$ to the center of the disk. This enables us to express the ratio by the simplified expression
\begin{equation}
R_{j,n} \;=\; \frac{\psi_{\mu,1}(w(z_{j_n}))}{\psi_{\mu,1}(0)} \;=\;\psi_{\mu,1}\bigl(w_n\bigr)\,,
\end{equation}
where the complex numbers $w_n=w(z_{j_n})$ are once again the positions of the  nearest neighbors in the central cell. As depicted in the right panel of Fig.~\ref{fig:taucalc}, they are uniformly distributed on a circle with radius $h$ defined in (\ref{RadiusH}). Defining the complex phase
\begin{equation}
Z=e^{2\pi i/q}
\end{equation}
we can represent them as $w_n = hZ^{\tau_j+n}$, implying that
\begin{equation}
\label{RnExplicit2}
R_{j,n} \;:=\;
\psi_{\mu,1}(hZ^{\tau_j+n})
\;=\;\left( \frac{\bigl|1-h Z^{\tau_j+n}\bigr|^2}{1-h^2} \right)^{\mu}\,.
\end{equation}
Assuming that the correction factors $\chi_j$ depend exclusively on the local inclination, i.e., $\chi_j = \chi(\tau_j)$, the eigenvalue problem (\ref{EigenvalueProblem1b}) turns into the following equation for the correction function $\chi(\tau)$:
\begin{equation}
\label{EigenvalueProblemCorrectionFunction}
\sum_{n=0}^{q-1}R_{j,n} \chi(\tau_{j_n}) \;=\; \bigl(q-\mathcal N \Lambda_\mu\bigr) \chi(\tau_j)\,.
\end{equation}
This equation is supposed to hold at all vertices $j$ of the infinite lattice. Given that the corresponding inclinations $\tau_j$ appear to densely fill the interval $[0,1)$ across the lattice, we can therefore consider just an arbitrary value $\tau\in[0,1)$ of some vertex as given and then compute the corresponding inclinations $\sigma_n(\tau)$ of its $q$ nearest neighbors, enumerated by $n=0,\ldots,q-1$. 

\begin{figure}
\includegraphics[width=85mm]{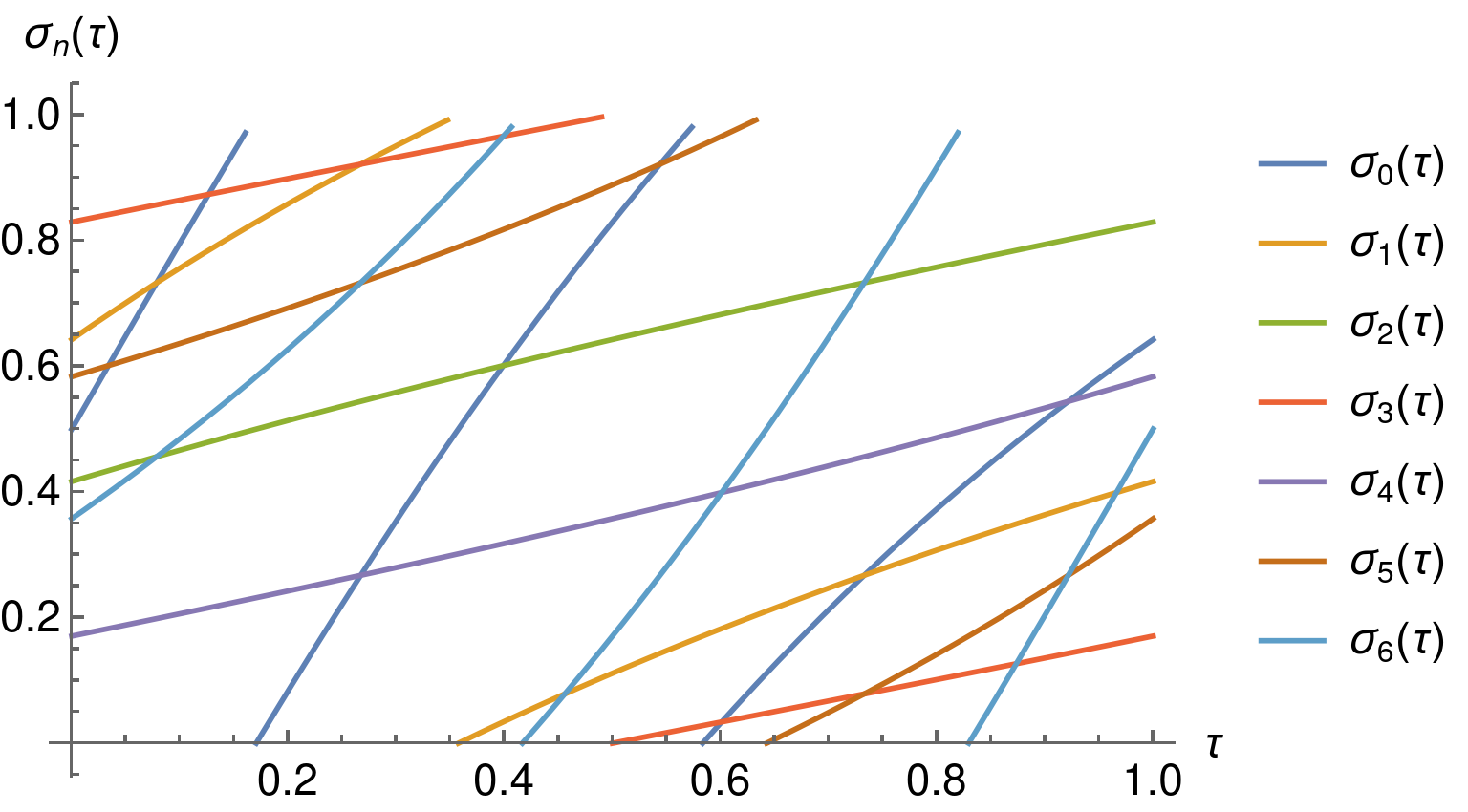}
\caption{Inclination $\sigma_n(\tau)$ of the $n^\text{th}$ neighbor of a vertex with inclination $\tau$ on a $\{3,7\}$ lattice.
\label{supfig:neighbor-inclination}}
\end{figure}

The neighboring inclinations $\sigma_n(\tau)$ can be determined as follows. As described above, we first use again the isometry (\ref{SpecialIsometry}) to map the cell to the center. Then, selecting a nearest neighbor at $h Z^{\tau+n}$, we map this neighbor in a second step isometrically to the center while keeping the source at $b=1$ fixed. The corresponding isometry reads
\begin{equation}
\tilde w(z) = \frac{(Z^{\tau+n}-h)(h Z^{\tau+n}-z)}{(hZ^{\tau+n}-1)(Z^{\tau+n}-hz)}\,. 
\end{equation}
The inclination $\sigma_n(\tau)$ of the $n^{\mathrm{th}}$ neighbor is then determined by the angular argument of $\tilde w(0)$, leading to the expression
\begin{equation}
\label{SigmaDef}
\sigma_n(\tau) \;=\; \frac{1}{2\pi} \arg \left[ \Bigl( \frac{h-Z^{\tau+n}}{1-h Z^{\tau+n}} \Bigr)^q \right] \in [0,1)\,.
\end{equation}
This provides $q$ different functions $\sigma_0(\tau),\ldots,\sigma_{q-1}(\tau)$. For given inclination $\tau$ at a some vertex, these functions allow us to compute the inclination of the nearest neighbors. An example of these functions in case of a $\{3,7\}$-lattice is shown in Fig.~\ref{supfig:neighbor-inclination}.

Rewriting (\ref{RnExplicit2}) as
\begin{equation}
\label{RnExplicit}
R_n(\tau) \;=\; \biggl( \frac{\bigl|1-h Z^{\tau+n}\bigr|^2}{1-h^2} \biggr)^{\mu}
\end{equation}
and replacing $\tau_j\to\tau$ and $\tau_{j_n}\to\sigma(\tau)$ in Eq.~(\ref{EigenvalueProblemCorrectionFunction}) we are led to the final equation
\begin{equation}
\label{DifferenceEquationAppendix}
\sum_{n=0}^{q-1} R_n(\tau) \, \chi\bigl(\sigma_{n}(\tau)\bigr) \;=\; \bigl(q-\mathcal N \Lambda_\mu\bigr) \chi(\tau)\,.
\end{equation}
This equation is expected to hold for all $\tau \in [0,1)$. It relates the function $\chi$ evaluated at $\tau$ linearly with itself evaluated at $q$ different inclinations $\sigma_0(\tau),\ldots,\sigma_{q-1}(\tau)$. Thus, it can be considered as a linear finite-difference equation for the correction function $\chi(\tau)$.

%=====================================================================
\section{Numerical Iteration Scheme for the Correction Function}
\label{AppendixNumericalScheme}
%=====================================================================

The difference equation~(\ref{DifferenceEquationAppendix}) can be regarded as an eigenvalue problem for $\chi(\tau)$, and it turns out that for $\mu\geq0$ the eigenfunctions with the largest eigenvalue correspond to the plane-wave solutions we aim to determine. This suggests a simple numerical iteration scheme. For this, we first divide the domain of the inclination $\tau\in[0,1)$ into $T$ equidistant bins labeled by the index $t=0,\ldots,T-1$, thereby discretizing the correction function $\chi(\tau)$ by a vector $\vec X$ with the components $X_t \simeq \chi(t/T)$. Setting
\begin{align}
R_{n,t}    &:= R_n(t/T)\\
s_{n,t} &:= \lfloor T \sigma_n(t/T)\rfloor
\end{align}
and initializing $X^{[0]}_t:=1$, we iterate $\vec X^{[k]}\mapsto \vec X^{[k+1]}$ via
\begin{align}
&Y_t \;:=\; \sum_{n=0}^{q-1} R_{n,t} X^{[k]}_{s_{n,t}}  \quad (t=0\ldots,T-1)
\\
&\eta \;:=\; \frac{1}{T}\sum_{t=0}^{T-1} |Y_t|
\\
&\vec X^{[k+1]} \;:=\; {\vec Y}/{\eta}
\end{align}
in a loop until the eigenvector $\vec X^{[k]}$ becomes stationary. The corresponding eigenvalue can then be estimated by 
\begin{equation}
\Lambda_\mu \approx \frac{1}{\mathcal N}(q-\eta)\,.
\end{equation}
For $\mu>0$ this iteration procedure converges quickly, while for $\mu<0$ the convergence is much slower. Note that this procedure determines the largest eigenvalue and that we implicitly assume that the plane-wave solution we are interested in just corresponds to the largest eigenvalue.

%=====================================================================
\section{Fourier Expansion}
\label{AppendixFourierExpansion}
%=====================================================================
%
As explained above, our numerical results strongly suggest that it is advantageous to express the correction function $\chi(\tau)$ as a Fourier series:
\begin{equation}
\label{FourierAnsatz2}
\chi(\tau) \;=\; \sum_{k=-\infty}^{+\infty} \gamma_k \, Z^{q\tau k}
\,,
\end{equation}
where $Z=e^{2\pi i/q}$ and $\gamma_k\in\mathbb C$ are certain Fourier coefficients yet to be determined. Since eigenvectors are only determined up to a factor, we are free to set $\gamma_0:=1$. We are now going to derive an eigenvalue equation that allows us to determine the coefficients $\{\gamma_k\}$.

Starting point is the difference equation~(\ref{DifferenceEquationAppendix}).
To simplify the notations, let us introduce the abbreviation
\begin{equation}
Y_n := 1-h  Z ^{\tau+n}
\end{equation}
and its complex conjugate $\bar Y_n := 1-h  Z ^{-\tau-n}$ with $h$ defined in~(\ref{RadiusH}). This enables us to rewrite Eqs.~(\ref{SigmaDef}) and~(\ref{RnExplicit}) by
\begin{align}
Z^{q\sigma_{n}(\tau)} \;&=\;  \Bigl( \frac{h-Z^{\tau+n}}{1-h Z^{\tau+n}} \Bigr)^q
\;=\; \frac{(-1)^{q}Z^{q\tau}\bar Y_n^q}{Y_n^q}
\,,\\
R_n(\tau) \;&=\;\biggl( \frac{\bigl|1-h Z^{\tau+n}\bigr|^2}{1-h^2} \biggr)^{\mu}
\;=\; \frac{Y^\mu_n \bar Y^\mu_n}{(1-h^2)^{\mu}} \,,
\end{align}
where we used $Z^{nq}=1$. Inserting the ansatz (\ref{FourierAnsatz2}), the difference equation~(\ref{DifferenceEquationAppendix}) can then be reformulated as
\begin{equation}
\begin{split}
\sum_{n=0}^{q-1}
\frac{Y^\mu_n \bar Y^\mu_n}{(1-h^2)^{\mu}}
\sum_{k=-\infty}^{+\infty} \gamma_k \frac{(-1)^{qk}Z^{q\tau k}\bar Y_n^{qk}}{Y_n^{qk}}\\
\;=\; \bigl(q-\mathcal N \Lambda_\mu\bigr)
\sum_{k=-\infty}^{+\infty} \gamma_k Z^{q\tau k}\,.
\end{split}
\end{equation}
Organizing the terms yields:
\begin{equation}
\begin{split}
\label{SEquation}
&\frac{1}{(1-h^2)^{\mu}}
\sum_{k=-\infty}^{+\infty} \gamma_k
(-1)^{qk}Z^{q\tau k}
\underbrace{\sum_{n=0}^{q-1}
Y_n^{\mu-qk}
\bar Y_n^{\mu+qk}}_{=: S_k}
\\[-6mm]
&\;=\; \bigl(q-\mathcal N \Lambda_\mu\bigr)
\sum_{k=-\infty}^{+\infty} \gamma_k Z^{q\tau k}\,.
\end{split}
\end{equation}
We first evaluate the part $S_k$ marked by the curly bracket using the general relation
\begin{equation}
(1+a)^\nu\;=\; \sum_{l=0}^\infty \binom \nu l a^l
\end{equation}
where $\binom a b = \frac{\Gamma(a+1)}{\Gamma(b+1)\Gamma(a-b+1)}$. 
The resulting expression reads
\begin{equation}
\begin{split}
S_k   &\;=\; \sum_{n=0}^{q-1} Y^{\mu-qk}_{n}\,\bar Y^{\mu+qk}_{n}
\\
&\;=\;
\sum_{l=0}^\infty \binom {\mu-qk} l
\sum_{m=0}^\infty \binom {\mu+qk} m 
\\
& \hspace{10mm} \times
 (-h)^{l+m}Z^{\tau(l-m)}\sum_{n=0}^{q-1}Z^{n(l-m)}
 \,.
\end{split}
\end{equation}
Given that $Z=e^{2\pi i/q}$, the last sum yields:
\begin{equation}
\sum_{n=0}^{q-1}Z^{n(l-m)} \;=\;
\begin{cases}
q & \hspace{3mm}\text{ if  } (l-m) \text{  is a multiple of } q \\
0 & \hspace{3mm}\text{ otherwise.}
\end{cases}
\end{equation}
The expression for $S_k$ therefore decomposes into three parts, namely, for $m=l$, for $m=l+qs$, and for $l=m+qs$ with $s \in \mathbb N$:
\begin{equation}
\begin{split}
S_k &= q \sum_{l=0}^\infty \binom {\mu-qk} l \binom {\mu+qk} l (-h)^{2l} \\
&+ q \sum_{s=1}^\infty \, \sum_{l=0}^\infty \binom {\mu-qk} l  \binom {\mu+qk} {l+qs} (-h)^{2l+qs}\, Z^{-\tau qs}\\
&+ q \sum_{s=1}^\infty \, \sum_{m=0}^\infty \binom {\mu-qk} {m+qs} \binom {\mu+qk} m (-h)^{2m+qs}\, Z^{\tau qs}
\end{split}
\end{equation}
It turns out that these sums can be expressed in terms of generalized hypergeometric functions
\begin{equation}
\begin{split}
S_k &= q\,\,
F_{-qk,qk}
\\
&+ q \sum_{s=1}^\infty \,\binom {\mu+qk} {qs}\,
F_{qk,qs-qk}
(-h)^{qs} Z^{-\tau qs}
\\
&+ q \sum_{s=1}^\infty \,\binom {\mu-qk} {qs}\,
F_{-qk,qs+qk}
(-h)^{qs} Z^{\tau qs}
\end{split}
\end{equation}
where we used again the abbreviation
\begin{equation}
\label{Abbreviation2}
F_{a,b}={}_2F_1\bigl(a-\mu,\,b-\mu;\,1+a+b;\,h^2\bigr)\,.
\end{equation}
Reinserting $S_k$ back into Eq.~(\ref{SEquation}) and comparing the coefficients in the powers of $Z$ on both sides, it is straight-forward to show that the Fourier coefficients comply with the eigenvalue problem
\begin{equation}
\label{GammaEVP}
\sum_{k=-\infty}^{\infty} B_{j,k}\,\gamma_k \;=\; (q-\mathcal N \Lambda_\mu) \gamma_j
\end{equation}
for all $j \in \mathbb Z$. Here the infinite-dimensional matrix $B$ with matrix elements $B_{j,k}$ is defined in terms of generalized hypergeometric functions as:
\begin{equation}
\label{matrixB}
B_{j,k} \;=\; \frac{q \,(-1)^{qj}\,h^{q|j-k|}}{(1-h^2)^\mu}
\begin{cases}
\binom{\mu-qk}{qj-qk}\,
F_{qj,-qk}
%\,{}_2F_1\bigl(-\mu+qj,\,-\mu-qk;\,1+qj-qk;\,h^2\bigr)
& \text{ if } j\geq k\\
\binom{\mu+qk}{qk-qj}\,
F_{-qj,qk}
%\,{}_2F_1\bigl(-\mu-qj,\,-\mu+qk;\,1+qk-qj;\,h^2\bigr)
& \text{ if } j<k
\end{cases}\,.
\end{equation}
Equivalently, we can rewrite this matrix in terms of generalized Jacobi functions $P_n^{\alpha,\beta}(z)$ as
\begin{small}\begin{equation}
\label{matrixB2}
B_{j,k} = \frac{q \,(-1)^{qj}\,h^{q|j-k|}}{(1-h^2)^\mu}
\begin{cases}
P_{\mu-qj}^{q(j-k),-1-2\mu}(1-2h^2) & \text{ if } j\geq k\\
P_{\mu+qj}^{q(k-j),-1-2\mu}(1-2h^2) & \text{ if } j<k
\end{cases}\,.
\end{equation}\end{small}
Note that this matrix is symmetric under reflections
\begin{equation}
B_{j,k}=B_{-j,-k}.
\end{equation}
Thus, the eigenvectors of (\ref{GammaEVP}) are either symmetric or antisymmetric ($\gamma_j=\pm \gamma_{-j}$), corresponding to real-valued or purely imaginary correction functions. Given our interest in real-valued symmetric solutions with $\gamma_j= \gamma_{-j}$, we can further condense the eigenvalue problem to:
\begin{equation}
\label{GammaEVPSymmetric}
\sum_{k=0}^{\infty} C_{j,k}\,\gamma_k \;=\; (q-\mathcal N \Lambda) \gamma_j
\end{equation}
for $j=0,1,2,\ldots$ with
\begin{equation}
C_{j,k} :=
\begin{cases}
B_{j,0} & \text{ if } k=0 \\
B_{j,k}+B_{j,-k} & \text{ if } k>0
\end{cases}
\end{equation}

%=====================================================================
\section{Exact Solution for $\mu=q,q+1,\ldots,2q-1$}
\label{AppendixExactSolution}
%=====================================================================
%
As an example of a nontrivial exact solution, let us study the second group of integer values in the range
\begin{equation}
\mu=q,q+1,\ldots,2q-1\,.
\end{equation}
For these values we have $s=1$, meaning that a $3\times 3$ block matrix in the center of $B$ decouples and can be diagonalized exactly. Here, we find the following explicit exact solution
\begin{equation}
\Psi_{\mu,b}(z_j) \;=\; \, \Bigl( 1+2\gamma_1\cos(2\pi \tau_j) \Bigr) \,\psi_{\mu,b}(z_j)
\end{equation}
Here the Fourier coefficients are given by $\gamma_0=1$ and
\begin{equation}
\gamma_1 =\frac{2  h^q \binom{\mu }{q} \,
F_{q,0}
}
{
(-1)^q\Bigl( F_{0,0}
+\sqrt{W}\Bigr)
- h^{2 q} \binom{q+\mu }{2 q} \,
F_{q,q}
-\,
F_{q,-q}
}
\,.
\end{equation}
\begin{widetext}
\noindent
The corresponding eigenvalue reads
\begin{equation}
\mathcal N \Lambda^{[s=1]}_\mu \;=\; q \,-\,
\frac{q}{2\left(1-h^2\right)^{\mu } }  \Biggl[
(-1)^q h^{2 q} \binom{q+\mu }{2 q} \,
F_{q,q}
\,+\,F_{0,0}
 +(-1)^q  \,
F_{q,-q}
+\sqrt{W}\Biggr]\,,
\end{equation}
where
\begin{small}
\begin{equation}
\begin{split}
W \;=\; & \Biggl[ (-1)^q F_{0,0}+F_{-q,q}
+h^{2q}\binom{q+\mu}{2q}F_{q,q}\Biggr] ^2 
+  8 (-1)^q h^{2q} \binom{\mu}{q} \binom{q+\mu}{q} F_{q,0}^2  - 4 (-1)^q F_{0,0} \Biggl[ F_{-q,q} + h^{2q} \binom{q+\mu}{2q} F_{q,q}\Biggr]
\end{split}
\end{equation}
\end{small}

Note that this nontrivial solution is exact across the entire infinite hyperbolic lattice.

\end{widetext}

%=====================================================================
\bibliography{library}
%=====================================================================

\end{document}